\documentclass{elsart}

\usepackage{graphicx,natbib,amssymb,lineno,hyperref}
\usepackage{lscape}

\def\astrobj#1{#1}

\newcommand\apj{{ApJ}}%
\newcommand\aj{{AJ}}%

\begin{document}

\begin{frontmatter}
\title{First analysis of eight Algol-type systems: V537~And, GS~Boo, AM~CrB, V1298~Her, EL~Lyn, FW~Per, RU~Tri, and WW~Tri}

\author{P. Zasche}
\ead{zasche@sirrah.troja.mff.cuni.cz}

\address{Astronomical Institute, Faculty of Mathematics and Physics,
 Charles University in Prague, CZ-180 00 Praha 8, V Hole\v{s}ovi\v{c}k\'ach 2, Czech Republic}

\begin{abstract}
Analyzing available photometry from the Super WASP and other databases, we performed the very first
light curve analysis of eight eclipsing binary systems \astrobj{V537~And}, \astrobj{GS~Boo},
\astrobj{AM~CrB}, \astrobj{V1298~Her}, \astrobj{EL~Lyn}, \astrobj{FW~Per}, \astrobj{RU~Tri}, and
\astrobj{WW~Tri}. All of these systems were found to be detached ones of Algol-type, having the
orbital periods of the order of days. 722 new times of minima for these binaries were derived and
presented, trying to identify the period variations caused by the third bodies in these systems.
\end{abstract}


\end{frontmatter}

\section{Introduction}

The eclipsing binaries provide us with an excellent method how to derive the basic physical
properties of the two eclipsing components (their radii, masses, temperatures). Moreover, they can
also serve as independent distance indicators, one can study the dynamical evolution of the orbits,
test the stellar structure models, or discover additional components in these systems (see e.g.
\citealt{2006Ap&SS.304....5G}). Due to these reasons and availability of the photometric
observations for some of these systems, we decided to carry out the first analysis for a few
eclipsing binaries which were never been studied before.

The analysis of the light curves (hereafter LC) became almost a routine task thanks to the programs
like {\sc PHOEBE} \citep{Prsa2005}. Also the photometric data are very easy to be obtained due to
long-term monitoring surveys covering a large fraction of the sky -- like NSVS
\citep{2004AJ....127.2436W}, ASAS \citep{2002AcA....52..397P}, Super WASP
\citep{2006PASP..118.1407P}, and others.

\section{Analysis}

The target selection for this paper was rather straightforward. We have chosen only these systems,
which are known to be eclipsing variables, their orbital period is known, has never been analysed
before and have enough photometric data points for an analysis. Thanks to the good time coverage
provided by the Super WASP survey we used this database for the whole analysis. All of the analysed
systems are the northern-hemisphere stars (DE $>$ 20$^\circ$) of moderate brightness (9.5~mag $<$ V
$<$ 13~mag) and with the orbital periods ranging from 0.6 to 3.3 days.

For the light curve analysis the {\sc PHOEBE} program \citep{Prsa2005} was used, which is based on
the algorithm by \cite{Wilson1971}. None of the selected stars was ever observed spectroscopically,
hence some of the parameters have to be fixed during the light curve solution. At first, the
"Detached binary" mode (in Wilson \& Devinney mode 2) was assumed for computing. The value of the
mass ratio $q$ was set to 1. The limb-darkening coefficients were interpolated from van~Hamme's
tables (see \citealt{vanHamme1993}), the linear cosine law was used. The values of the gravity
brightening and bolometric albedo coefficients were set at their suggested values for convective or
radiative atmospheres (see \citealt{Lucy1968}). 
Therefore, the quantities which could be directly calculated from the light curve are the
following: the relative luminosities $L_i$, the temperature of the secondary $T_2$, the inclination
$i$, and the Kopal's modified potentials $\Omega_1$ and $\Omega_2$. The synchronicity parameters
$F_1$ and $F_2$ were also fixed at values of 1. The value of the third light $L_3$ was also
computed if a non-negligible value resulted from the fitting process. And finally, the linear
ephemerides were calculated using the available minima times for a particular system.

With the final LC analysis, we also derived many times of minima for a particular system, using a
method presented in \cite{ZascheSMC2014}. The template of the LC is used to fit the photometric
data from the Super WASP as well as from other surveys, resulting in a set of minima times, which
can be used for a subsequent period analysis. The already published observations were also used for
the analysis, mostly taken from the $(O-C)$ gateway\footnote{http://var.astro.cz/ocgate}
\citep{2006OEJV...23...13P}.

\section{The individual systems}

\subsection{V537 And}

The system V537 And (= GSC 02814-01959, V=11.2~mag) is relatively neglected eclipsing binary of
Algol-type. It was first mentioned by \cite{2008PZP.....8...41K}, who analysed the NSVS photometry
and found an orbital period of about 0.9~days and relatively deep eclipses of about 0.3~mag.
However, since then no detailed analysis of this target was performed, only a few times of minima
were published to better constrain the orbital period \citep{OEJV160}.

\begin{figure}
 \includegraphics[width=\textwidth]{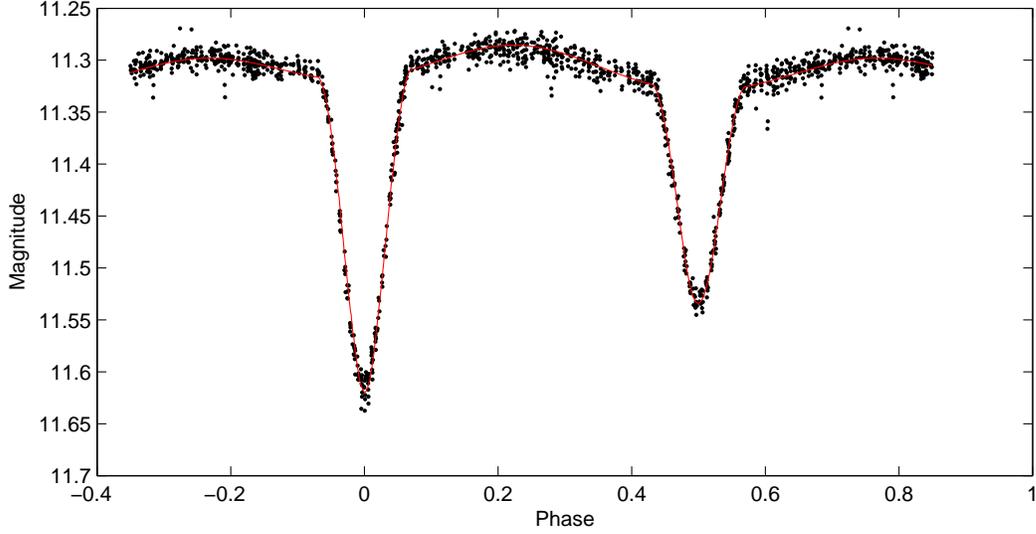}
 \caption{Light curve analysis of V537 And, based on the Super WASP photometry.}
 \label{FigV537AndLC}
\end{figure}

We extracted the Super WASP photometry of the star for the LC analysis. However, only a small
fraction of the data points were used for the LC modelling (these ones with better precision and
obtained during a shorter time span). For the light curve fitting process one of the most crucial
parameters is the value of the primary temperature $T_1$, which is kept fixed during the whole
fitting. Due to the fact that the star was included into the Tycho survey onboard the Hipparcos
satellite, as well as observed during the 2MASS and NOMAD surveys, \cite{2010PASP..122.1437P} used
all of these data to roughly estimate the spectral type of the system. This resulted in G0V, which
is the only available spectral estimation of V537~And. Therefore, we fixed the primary temperature
at a value of $T_1=5900$~K, in agreement with its spectral type (see e.g.
\citealt{1988BAICz..39..329H}). The {\sc PHOEBE} program was used and the final fit is presented in
Fig.\ref{FigV537AndLC}. As one can see, the magnitudes during both maxima are different, hence also
the hypothesis of a photospheric spot on the surface of the primary component was used. The final
LC parameters are presented in Table \ref{TableLC}. The secondary component is a bit cooler and
smaller, while no third light was detected.

\begin{figure}
 \includegraphics[width=\textwidth]{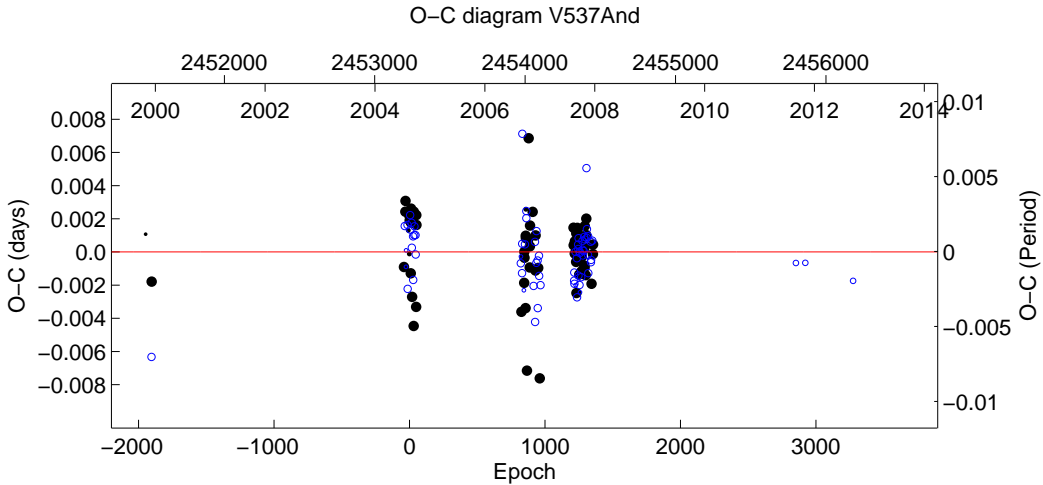}
 \caption{O-C diagram of times of minima derived from available photometry for V537~And. The black points stand for the
 primary minima, while the blue open circles stand for the secondary ones. The larger the symbol, the higher the weight.}
 \label{FigV537AndOC}
\end{figure}

We also used the LC template for deriving the times of minima from the available photometry (Super
WASP and NSVS). All of these data points are stored in the Table \ref{TableMin}.  In total, 137 new
times of minima were derived. Together with four already published minima, these observations can
tell us something about the long-term evolution of the orbit (see Fig. \ref{FigV537AndOC}).
However, no additional variation is visible on these data points (no third-body modulation of the
orbit), which is in agreement with no third light detected from the LC solution. However, one can
speculate about the very first and the very last data points plotted, and maybe a parabolic fit
would give us a better solution (which can be attributed to the mass transfer between the
components). However, a longer time span is needed for this to be definitely solved. The linear
ephemerides as resulted from the analysis of minima times led to the values $JD_0$ and $P$ as
presented in Table \ref{TableLC}.

\begin{table*}[t]
 \tiny
 \caption{The light-curve parameters as derived from our analysis.}
 \label{TableLC} \centering
\begin{tabular}{ c c c c c c c}
\hline \hline
 Parameter     &    V537 And              &    GS Boo               &      AM~CrB               &    V1298~Her              \\ \hline
 $JD_0-2400000$& 53231.6375 $\pm$ 0.0028  & 53128.4529 $\pm$ 0.0025 & 51242.8130 $\pm$ 0.0042   & 53726.4961 $\pm$ 0.0068   \\
 $P$ [d]       &0.9008483 $\pm$ 0.0000012 &1.2568178 $\pm$ 0.0000017& 0.7036534 $\pm$ 0.0000010 & 2.1446937 $\pm$ 0.0000030 \\
 $q$ [$10^{-10}$d]&   --                  & -36.0 $\pm$ 2.2         &   --                      &  --                       \\
 $i$ [deg]     & 77.9 $\pm$ 1.2           & 76.9 $\pm$ 0.8          &  77.1 $\pm$ 1.8           &  83.6 $\pm$ 0.7           \\
 $T_1$ [K]     & 5900 (fixed)             & 6400 (fixed)            &  7000 (fixed)             &  6100 (fixed)             \\
 $T_2$ [K]     & 5310 $\pm$ 280           & 4410 $\pm$ 260          &  3810 $\pm$ 430           &  5410 $\pm$ 290           \\
 $\Omega_1$    & 4.940 $\pm$ 0.019        & 5.683 $\pm$ 0.028       & 4.134 $\pm$ 0.013         &  7.102 $\pm$ 0.032        \\
 $\Omega_2$    & 5.921 $\pm$ 0.021        & 4.748 $\pm$ 0.021       & 4.534 $\pm$ 0.020         &  10.671 $\pm$ 0.060       \\
 $L_1$ [\%]    & 70.7 $\pm$ 1.1           & 85.4 $\pm$ 1.1          & 94.4 $\pm$ 1.2            &  80.6 $\pm$ 1.2           \\
 $L_2$ [\%]    & 29.3 $\pm$ 0.9           & 14.6 $\pm$ 0.5          &  4.1 $\pm$ 0.7            &  18.7 $\pm$ 0.9           \\
 $L_3$ [\%]    &  0.0 $\pm$ 0.0           &  0.0 $\pm$ 0.0          &  1.5 $\pm$ 0.6            &   0.7 $\pm$ 0.9           \\ \hline
   \multicolumn{1}{c}{Spots:}             &                         &                           &                           \\
 $b_1$ [deg]   &  13.7 $\pm$ 1.7          &  --                     &   --                      &   --                      \\
 $l_1$ [deg]   & 330.0 $\pm$ 12.0         &  --                     &   --                      &   --                      \\
 $r_1$ [deg]   &  13.2 $\pm$ 1.2          &  --                     &   --                      &   --                      \\
 $k_1$         &  1.34 $\pm$ 0.05         &  --                     &   --                      &   --                      \\ \hline
\end{tabular}
\end{table*}

\subsection{GS Boo}

The star GS Boo (=GSC 02565-00667, V=11.1~mag) is also rather seldom-investigated system. It was
discovered during the ROTSE survey \citep{2000AJ....119.1901A}, who presented incorrect period of
0.63~days, while the correct one is double, about 1.26~days. Later, only several papers with the
minima observations were published. No light curve nor spectroscopic analysis were performed.

\begin{figure}
 \includegraphics[width=\textwidth]{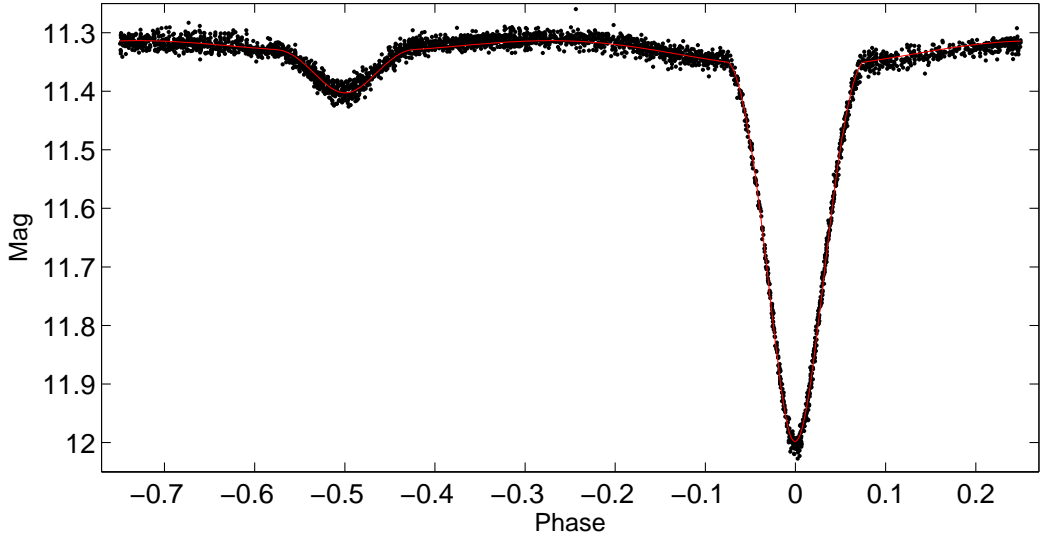}
 \caption{Light curve analysis of GS Boo, based on the Super WASP photometry.}
 \label{FigGSBooLC}
\end{figure}

Therefore, we used the Super WASP photometry to analyse the LC of GS~Boo. Despite a huge number of
data points (more than 18000), we used only a small portion of these data for a LC analysis. The
same situation as for V537~And also applies here, and \cite{2010PASP..122.1437P} gave the only
estimation about its spectral type. Hence, using an assumption of F6V component, we fixed the
primary temperature to the value of 6400~K for the whole fitting process. The result of the light
curve fit as provided by the {\sc PHOEBE} program is plotted in Fig. \ref{FigGSBooLC}, while the
parameters are given in Table \ref{TableLC}.

\begin{figure}
 \includegraphics[width=\textwidth]{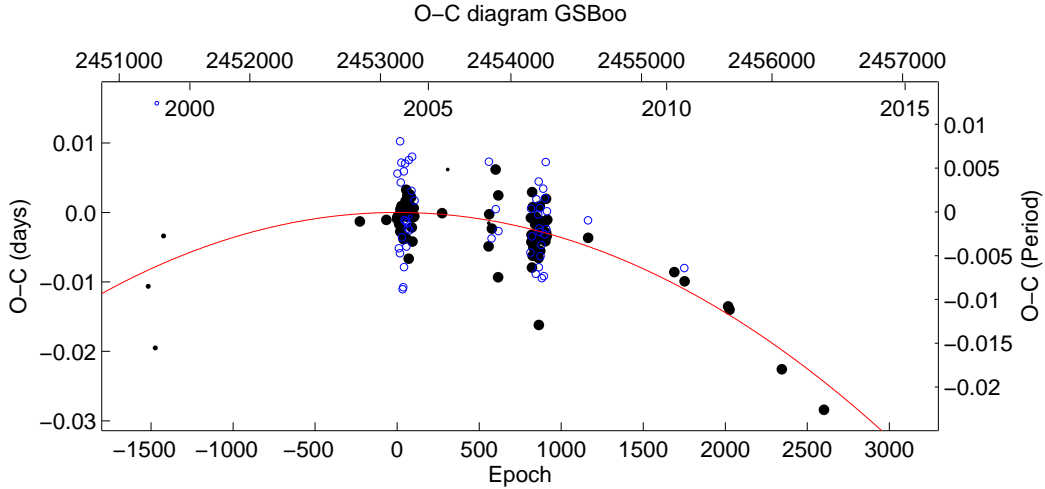}
 \caption{O-C diagram of times of minima for GS Boo.}
 \label{FigGSBooOC}
\end{figure}

The results of the minima fitting to the whole photometric data set are given in Table
\ref{TableMin}. As one can see from the $O-C$ diagram given in Fig. \ref{FigGSBooOC}, there is
evident a long steady decrease of the orbital period, hence the quadratic term for the ephemerides
was used (see the parabolic fit in Fig. \ref{FigGSBooOC}). Despite the fact the very last two data
points slightly deviate from the fit, any speculation about a possible third body is still rather
premature yet.

\subsection{AM~CrB}

AM~CrB (=GSC 02579-00069, V=12.6~mag) is an Algol-type eclipsing binary discovered during the ROTSE
survey \citep{2000AJ....119.1901A}. They also gave its correct orbital period of about 0.7~days,
but since then no other detailed analysis was performed.

\begin{figure}
 \includegraphics[width=\textwidth]{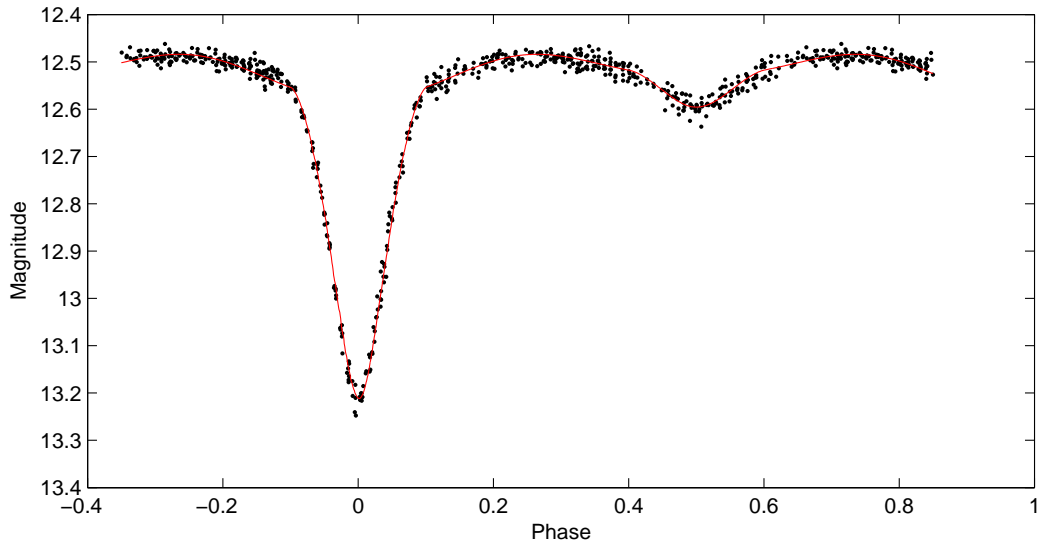}
 \caption{Light curve analysis of AM CrB, based on the Super WASP photometry.}
 \label{FigAMCrBLC}
\end{figure}

Due to limited information about the system, the spectral estimation by \cite{2010PASP..122.1437P}
was used (F0III), fixing the value of the temperature to 7000~K for the whole LC analysis. The
{\sc PHOEBE} program provides us with a solution presented in Fig. \ref{FigAMCrBLC}, while the LC
parameters are given in Table \ref{TableLC}. As one can see, the secondary component is rather
cooler and a bit smaller than the primary. A small fraction of the third light is on the lower limit
what can be detected via this method and only further more detailed analysis of much more
precise photometric data can confirm or refuse this hypothesis.

\begin{figure}
 \includegraphics[width=\textwidth]{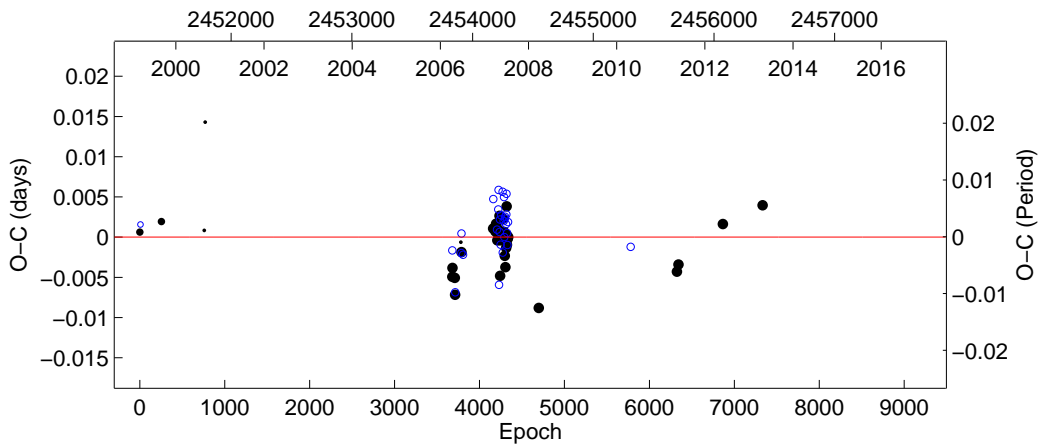}
 \caption{O-C diagram of times of minima for AM CrB.}
 \label{FigAMCrBOC}
\end{figure}

The $O-C$ diagram of the times of minima is plotted in Fig. \ref{FigAMCrBOC}, where the already
published data points (11 observations) are plotted together with our new times of minima as
derived from the Super WASP photometry (68 data points). The scatter of the new minima is rather
large, but mainly the minima after the year 2010 show some variation, which can possibly be
attributed to the third-body variation. New precise observations and minima timings during the
upcoming years would be very useful for discussing this effect and its possible confirmation.

\subsection{V1298~Her}

The eclipsing binary system V1298~Her (=GSC 02077-00730, V=9.8~mag) is the brightest system in our
sample. It was first mentioned as a variable by \cite{2007A&A...467..785N}, who also presented its
correct orbital period of about 2.14~days. The spectral type was derived as F8 by
\cite{1988A&AS...74..449R}. No other detailed analysis was performed.

\begin{figure}
 \includegraphics[width=\textwidth]{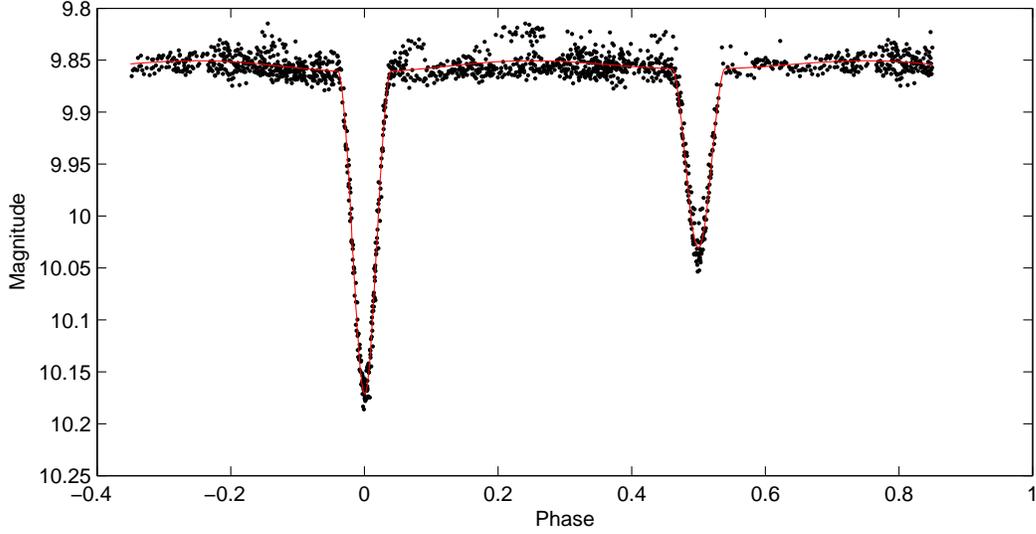}
 \caption{Light curve analysis of V1298 Her, based on the Super WASP photometry.}
 \label{FigV1298HerLC}
\end{figure}

The primary temperature was fixed at a value of 6100~K in agreement with its spectral type, and the
LC solution was found with the {\sc PHOEBE} program. The final fit is presented in Fig.
\ref{FigV1298HerLC}, and the parameters of the LC solution are given in Table \ref{TableLC}.
However, as one can see from the LC in Fig. \ref{FigV1298HerLC}, the shape of the LC changes over
the time interval and this intrinsic variability has certainly some influence on the LC solution
and precision of the LC parameters. Such variability is visible over many orbital revolutions of
the binary in the Super WASP data, but its origin remains an open question.

\begin{figure}
 \includegraphics[width=\textwidth]{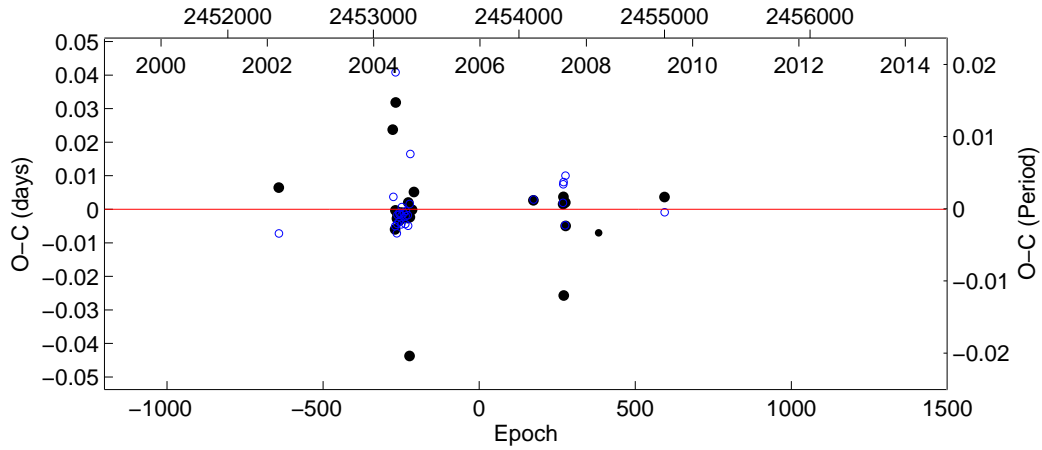}
 \caption{O-C diagram of times of minima for V1298 Her.}
 \label{FigV1298HerOC}
\end{figure}

The analysis of period of V1298 Her was based mainly on the Super WASP data, which yielded 62 times
of minima (see Table \ref{TableMin} and the final $O-C$ diagram in Fig. \ref{FigV1298HerOC}).
Together with a few data points derived from the ASAS photometry \citep{2002AcA....52..397P}, the
time span covered with the minima times ranges over eight years. However, no variation is visible
on these data points.

\begin{table*}[t]
 \tiny
 \caption{The light-curve parameters as derived from our analysis.}
 \label{TableLC2} \centering
\begin{tabular}{ c c c c c c c}
\hline \hline
 Parameter       &    EL~Lyn                &   FW~Per                  &    RU~Tri                 &   WW~Tri                 \\ \hline
 $JD_0-2400000$  & 55304.4386 $\pm$ 0.0022  & 54393.6419 $\pm$ 0.0036   & 53998.4392 $\pm$ 0.0066   & 53613.4883 $\pm$ 0.0056  \\
 $P$ [d]         & 0.6445384 $\pm$ 0.0000011& 0.7912319 $\pm$ 0.0000009 & 3.2685433 $\pm$ 0.0000050 & 1.7484409 $\pm$ 0.0000021\\
 $i$ [deg]       & 74.9 $\pm$ 0.6           & 89.9 $\pm$ 1.2            & 77.9 $\pm$ 1.2            & 85.1 $\pm$ 0.6           \\
 $T_1$ [K]       & 8800 (fixed)             & 6900 (fixed)              & 5600 (fixed)              & 7800 (fixed)             \\
 $T_2$ [K]       & 4920 $\pm$ 155           & 5060 $\pm$ 20             & 3980 $\pm$ 30             & 4700 $\pm$ 40            \\
 $\Omega_1$      & 4.204 $\pm$ 0.025        & 4.392 $\pm$ 0.013         & 4.424 $\pm$ 0.074         & 5.436 $\pm$ 0.019        \\
 $\Omega_2$      & 4.374 $\pm$ 0.072        & 4.760 $\pm$ 0.014         & 4.825 $\pm$ 0.057         & 7.612 $\pm$ 0.033        \\
 $L_1$ [\%]      & 80.3 $\pm$ 2.5           & 81.0 $\pm$ 1.4            & 83.7 $\pm$ 0.9            & 93.9 $\pm$ 0.9           \\
 $L_2$ [\%]      &  9.6 $\pm$ 0.8           & 18.9 $\pm$ 1.0            & 10.6 $\pm$ 1.4            &  6.1 $\pm$ 0.7           \\
 $L_3$ [\%]      & 10.1 $\pm$ 1.7           & 0.1 $\pm$ 0.2             &  5.7 $\pm$ 1.2            &  0.0 $\pm$ 0.0           \\ \hline
\end{tabular}
\end{table*}

\subsection{EL~Lyn}

EL Lyn (=GSC 02977-01179, V=12.6~mag) is another Algol-type eclipsing binary lacking of any
detailed analysis. The star was discovered as a variable by \cite{2005IBVS.5586....1O}, who gave
its correct orbital period of about 0.64~days, which makes it the shortest one in our sample. Later,
only a few minima observations were published, but other information is missing.

\begin{figure}
 \includegraphics[width=\textwidth]{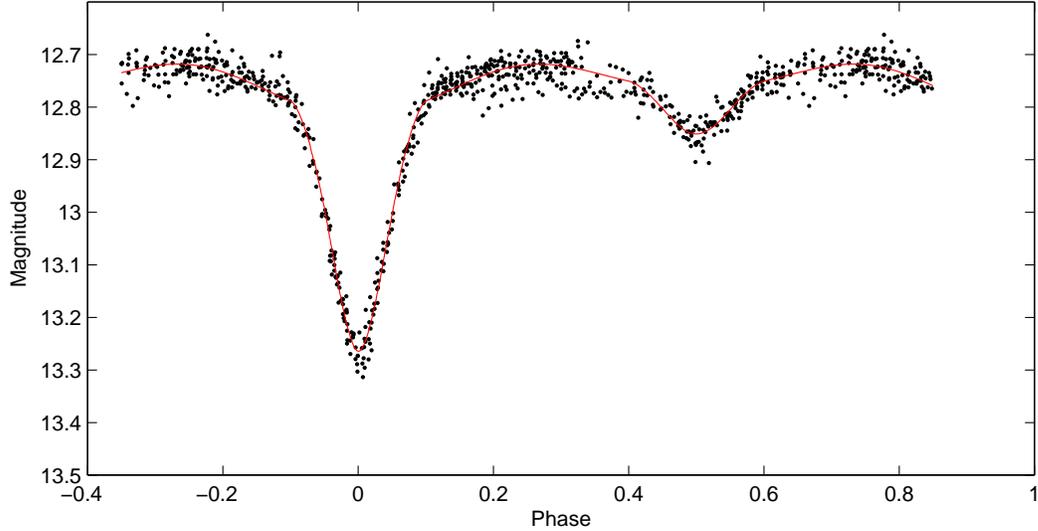}
 \caption{Light curve analysis of EL Lyn, based on the Super WASP photometry.}
 \label{FigELLynLC}
\end{figure}

Owing to having no information about its spectral type, only the photometric indices from the Tycho
$B-V=0.12$~mag, and $J-H= 0.358$~mag can be used to roughly estimate its type. Hence, we assumed
the spectral type of about A2, therefore the temperature of $T_1=8800$~K was kept fixed for the
whole LC analysis. The result of the LC fitting is plotted in Fig. \ref{FigELLynLC}, and the LC
parameters are given in Table \ref{TableLC2}. As one can see, the secondary component is
significantly cooler, but only mildly smaller than the primary. The level of the third light seems
to be non-negligible and one can hope to find the third body evident in upcoming years during some
more detailed photometric or spectroscopic analysis.

\begin{figure}
 \includegraphics[width=\textwidth]{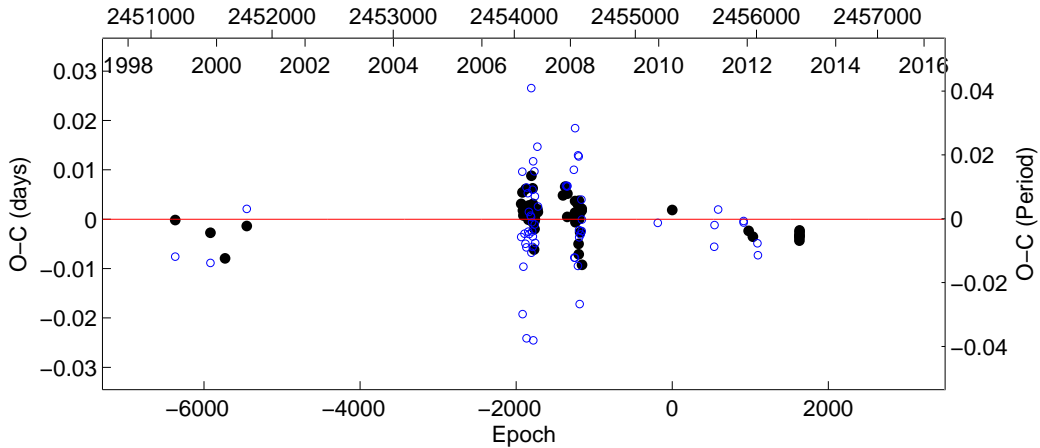}
 \caption{O-C diagram of times of minima for EL Lyn.}
 \label{FigELLynOC}
\end{figure}

The period analysis was done using the already published data (18 times of minima), while six new
data points were derived from the NSVS photometry and 104 from the Super WASP data. All of these
minima are given in the Table \ref{TableMin} and the final $O-C$ diagram is plotted in Fig.
\ref{FigELLynOC}. Regrettably, no additional variation is visible on the current data set and
longer time span is needed to be covered with observations.

\subsection{FW~Per}

The eclipsing system FW~Per (=GSC 03341-00406, V=12.5~mag) is an Algol-type eclipsing binary
discovered by \cite{1943AN....274...36H}. Despite its rather early discovery, the star was not
analysed neither photometrically, nor spectroscopically, and only a few publications with the
minima timings were published to date. It is the northernmost star in our sample, has the orbital
period of about 0.79~days, and shows rather deep eclipses.

\begin{figure}
 \includegraphics[width=\textwidth]{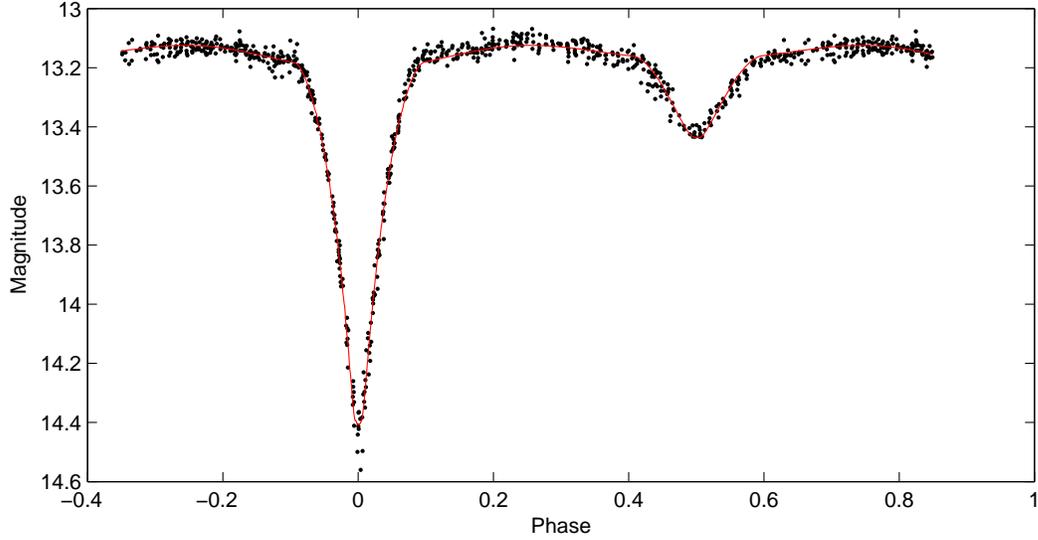}
 \caption{Light curve analysis of FW Per, based on the Super WASP photometry.}
 \label{FigFWPerLC}
\end{figure}

Using the Super WASP photometry, we analysed the light curve of FW~Per using the {\sc PHOEBE}
program. Owing to the fact that no spectral information is available, the photometric indices as
derived from the 2MASS photometry \citep{2006AJ....131.1163S} indicate that the star should be
earlier than F6, hence we fixed the primary temperature to the value of 6900~K for the whole LC fitting
process. With this assumption the LC was fitted, the final plot is given in Fig. \ref{FigFWPerLC},
and the parameters are written in Table \ref{TableLC2}.

\begin{figure}
 \includegraphics[width=\textwidth]{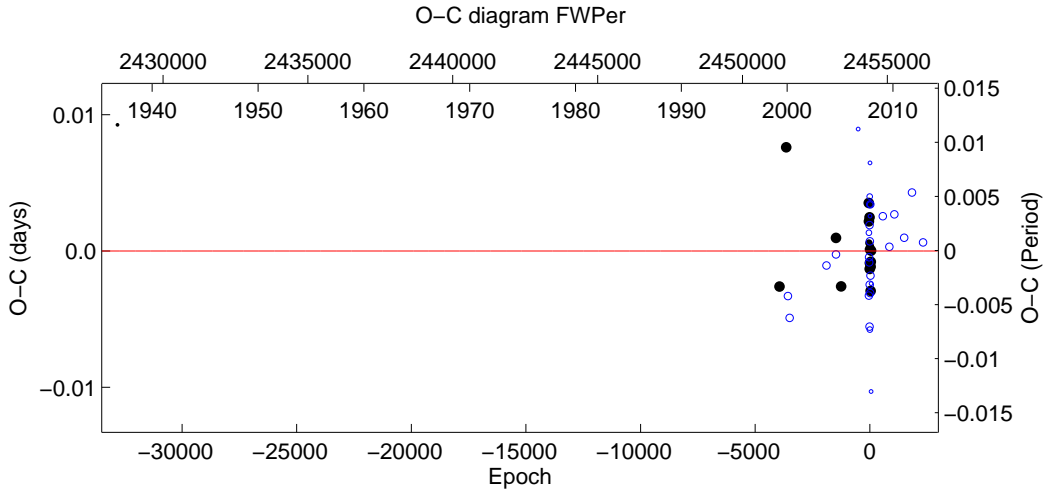}
 \caption{O-C diagram of times of minima for FW Per.}
 \label{FigFWPerOC}
\end{figure}

The analysis of period was done using the already published minima times, as well as with those
derived from the Super WASP data (36 new minima were derived). The final plot with the $O-C$
diagram is plotted in Fig. \ref{FigFWPerOC}, while no visible variation is evident on this plot.

\subsection{RU~Tri}

The star RU~Tri (=GSC 02316-00135, V=11.1~mag) was first mentioned as a variable by
\cite{1955VeBam..11....1S}. However, since then no detailed analysis was carried out. It has the
orbital period of about 3.3~days and shows relatively deep minima of about 0.6~mag (V filter).

\begin{figure}
 \includegraphics[width=\textwidth]{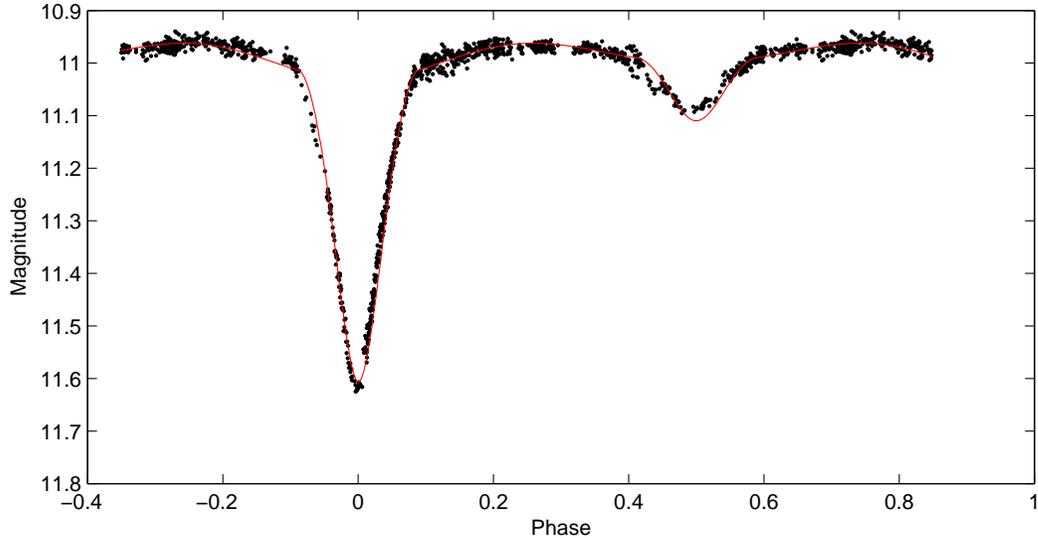}
 \caption{Light curve analysis of RU Tri, based on the Super WASP photometry.}
 \label{FigRUTriLC}
\end{figure}

Using the Super WASP photometry, we analysed the LC of the system. The spectral type was presented
as G0 by \cite{2006A&A...446..785M}, however later \cite{2010PASP..122.1437P} gave G5V type. Hence,
we use the latter value and fixed the primary temperature to the value of 5600~K. Moreover,
\cite{GCVS} noted also the close companion (22$^{\prime\prime}$ distant, 12~mag bright), but the
Washington double star catalogue \citep{WDS} does not include any such information, hence its
connection with the star RU~Tri is doubtful. Therefore, a light contamination from the close
component is expected due to the angular resolution of the Super WASP data
\citep{2010A&A...520L..10B}. The final LC fit is presented in Fig. \ref{FigRUTriLC}, while the
parameters are given in Table \ref{TableLC2}. The third light value as resulted from the LC
solution is surprisingly low for a close companion of such a brightness. We can only speculate that
only a fraction of its light enters the aperture of the Super WASP telescope. Another problematic
issue was some kind of additional intrinsic variability of the light curve as detected on the Super
WASP data. Nature of these variations still remains an open question, but it surely influence the
LC fit and its precision. If is this variation somehow connected with the close companion cannot
easily be solved with the current data.

\begin{figure}
 \includegraphics[width=\textwidth]{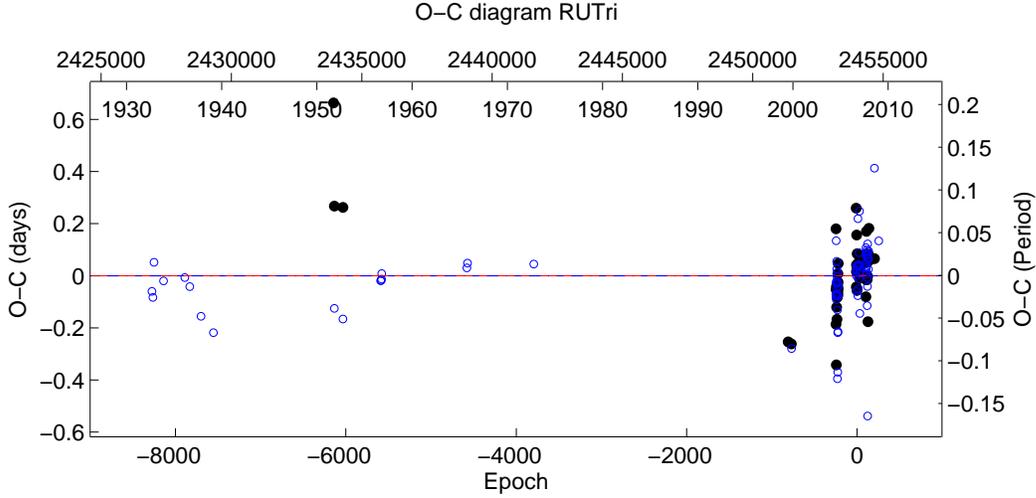}
 \caption{O-C diagram of times of minima for RU Tri.}
 \label{FigRUTriOC}
\end{figure}

On the other hand, also the times of minima were derived from the Super WASP and NSVS data. These
are stored in the Table \ref{TableMin}, and the final $O-C$ plot is given in Fig. \ref{FigRUTriOC}.
Despite rather large scatter of the data points, due to poor coverage any variation is evident.

\subsection{WW~Tri}

WW~Tri (=TYC 2322-796-1, V=12.1~mag) is an Algol-type eclipsing binary discovered by
\cite{1963IBVS...21....1W}. Its orbital period is of about 1.7~days, and shows rather deep eclipses
of about 0.6~mag. The spectral type was roughly estimated as A7V by \cite{2010PASP..122.1437P},
which is in good agreement with the $BVRI$ observations by \cite{2007yCat.2277....0S}. No detailed
analysis of the star was performed.

\begin{figure}
 \includegraphics[width=\textwidth]{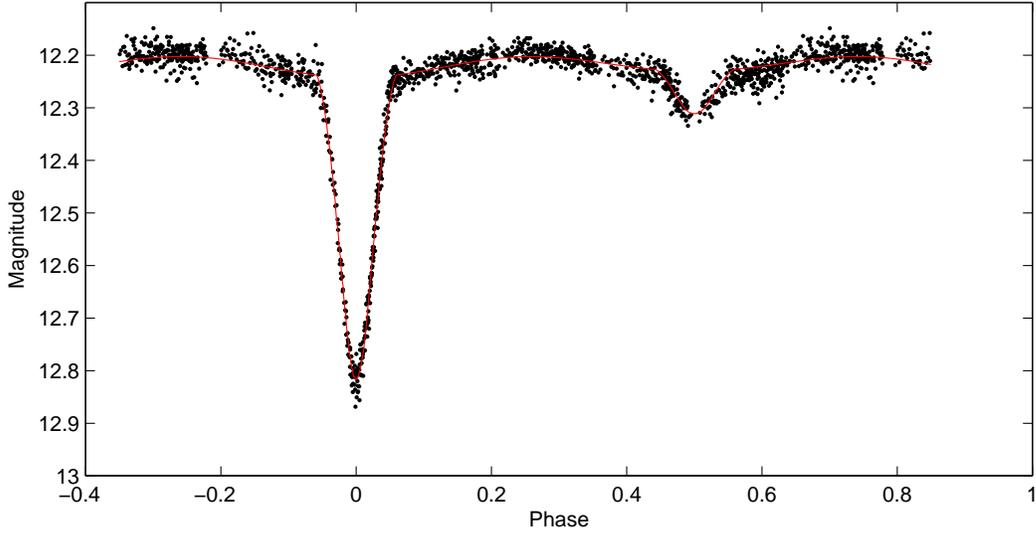}
 \caption{Light curve analysis of WW Tri, based on the Super WASP photometry.}
 \label{FigWWTriLC}
\end{figure}

Therefore, for the LC analysis the primary temperature of 7800~K was fixed. The Super WASP data
were used for the LC fitting in the {\sc PHOEBE} program. The final fit is presented in Fig.
\ref{FigWWTriLC}, and the parameters are given in Table \ref{TableLC2}. As one can see, the
secondary is significantly cooler and smaller, hence also its contribution to the total luminosity
of the system is only a few percent.

\begin{figure}
 \includegraphics[width=\textwidth]{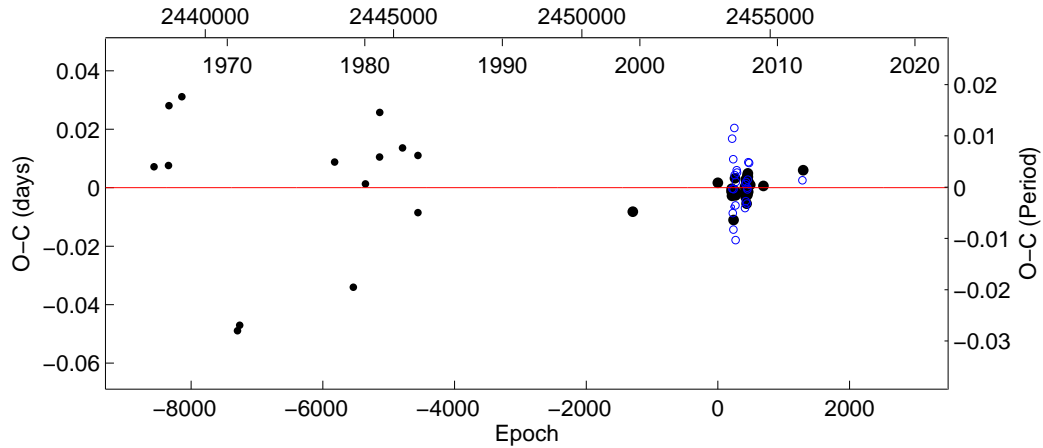}
 \caption{O-C diagram of times of minima for WW Tri.}
 \label{FigWWTriOC}
\end{figure}

The analysis of the period was carried out with the already published data points (20 minima),
together with our new data as derived from the Super WASP (63 minima). The final $O-C$ diagram is
plotted in Fig. \ref{FigWWTriOC}, but any visible variation is evident in the present data set.
More observations are needed during the next decades.

\section{Discussion and conclusions}

The first light curve analysis of eight Algol-type eclipsing binaries based on the Super WASP
photometric data led to several interesting results:

\begin{itemize}
  \item Confirming the well-known finding that for classical Algols the difference between the
  minima depths is in correlation with the temperature difference of the components.
  \item The surveys like Super WASP are suitable for this kind of analysis and the LC parameters
  as resulted from the analysis have only relatively small errors. Also the second-order effects
  like the third light contribution is detectable in these quality of data.
  \item The intrinsic variability is easily detectable in the Super WASP data due to the LC
  fitting and studying the residuals.
  \item Deriving the large set of minima timings for a prospective period analysis can be done
  relatively easily using the LC template.
  \item Using also some other surveys and the minima fitting procedure one can detect an additional
  variation in the $O-C$ diagram. But this method is more suitable for these systems, which have
  rather longer time span of observations for the third body perturbations to be discovered.
\end{itemize}

All of the presented systems are rather seldom-investigated and their follow-up observations using
spectroscopy would be of great benefit. Especially those ones, where some indication of additional
component in the system was found.


\section{Acknowledgments}
We thank the "ASAS", "NSVS", and "Super WASP" teams for making all of the observations easily
public available. This investigation was supported by the Czech Science Foundation grant no.
P209/10/0715, by the Research Program MSM 0021620860 of the Ministry of Education of Czech
Republic, and by the grant UNCE 12 of the Charles University in Prague. This research has made use
of the SIMBAD database, operated at CDS, Strasbourg, France, and of NASA's Astrophysics Data System
Bibliographic Services.

\begin{table}
 \tiny
 \caption{The heliocentric minima times used for the analysis.}
 \label{TableMin} \centering \scalebox{0.80}{
\begin{tabular}{ c c c c c c}
\hline \hline
     Star  &  HJD        & Error   & Type & Filter & Source    \\
           & 2400000+    & [days]  &      &        &           \\ \hline
 V537 And  & 53194.70023 & 0.00175 & Prim &  W  &  Super WASP  \\
 V537 And  & 53199.65736 & 0.00150 & Sec  &  W  &  Super WASP  \\
 V537 And  & 53203.71204 & 0.00169 & Prim &  W  &  Super WASP  \\
 V537 And  & 53204.61354 & 0.00105 & Prim &  W  &  Super WASP  \\
 V537 And  & 53207.76254 & 0.01098 & Sec  &  W  &  Super WASP  \\
 V537 And  & 53208.66439 & 0.01228 & Sec  &  W  &  Super WASP  \\
\hline
\end{tabular}}
\end{table}

\begin{table}
 \tiny
 \caption{The heliocentric minima times used for the analysis.}
 \label{TableMin} \centering \scalebox{0.80}{
\begin{tabular}{ c c c c c c}
\hline \hline
     Star  &  HJD        & Error   & Type & Filter & Source    \\
           & 2400000+    & [days]  &      &        &           \\ \hline
 V537 And  & 54102.30458 & 0.00212 & Sec  &  W  &  Super WASP  \\
 V537 And  & 54321.66479 & 0.00186 & Prim &  W  &  Super WASP  \\
 V537 And  & 54322.56459 & 0.00434 & Prim &  W  &  Super WASP  \\
 V537 And  & 54325.71542 & 0.00349 & Sec  &  W  &  Super WASP  \\
 V537 And  & 54326.61677 & 0.00220 & Sec  &  W  &  Super WASP  \\
 V537 And  & 54327.51694 & 0.00426 & Sec  &  W  &  Super WASP  \\
\hline
\end{tabular}}
\end{table}

\begin{table}
 \tiny
 \caption{The heliocentric minima times used for the analysis.}
 \label{TableMin} \centering \scalebox{0.80}{
\begin{tabular}{ c c c c c c}
\hline \hline
     Star  &  HJD        & Error   & Type & Filter & Source    \\
           & 2400000+    & [days]  &      &        &           \\ \hline
 GS Boo    & 53152.32972 & 0.00199 & Prim &  W  &  Super WASP  \\
 GS Boo    & 53153.58975 & 0.00113 & Prim &  W  &  Super WASP  \\
 GS Boo    & 53155.47883 & 0.00400 & Sec  &  W  &  Super WASP  \\
 GS Boo    & 53157.35934 & 0.00207 & Prim &  W  &  Super WASP  \\
 GS Boo    & 53158.61436 & 0.00170 & Prim &  W  &  Super WASP  \\
 GS Boo    & 53160.50894 & 0.00095 & Sec  &  W  &  Super WASP  \\
\hline
\end{tabular}}
\end{table}

\begin{table}
 \tiny
 \caption{The heliocentric minima times used for the analysis.}
 \label{TableMin} \centering \scalebox{0.80}{
\begin{tabular}{ c c c c c c}
\hline \hline
     Star  &  HJD        & Error   & Type & Filter & Source    \\
           & 2400000+    & [days]  &      &        &           \\ \hline
 GS Boo    & 54195.48584 & 0.00264 & Prim &  W  &  Super WASP  \\
 GS Boo    & 54206.79959 & 0.00662 & Prim &  W  &  Super WASP  \\
 GS Boo    & 54208.68754 & 0.00323 & Sec  &  W  &  Super WASP  \\
 GS Boo    & 54210.57153 & 0.00022 & Prim &  W  &  Super WASP  \\
 GS Boo    & 54211.82330 & 0.00115 & Prim &  W  &  Super WASP  \\
 GS Boo    & 54212.46273 & 0.00095 & Sec  &  W  &  Super WASP  \\
\hline
\end{tabular}}
\end{table}

\begin{table}
 \tiny
 \caption{The heliocentric minima times used for the analysis.}
 \label{TableMin} \centering \scalebox{0.80}{
\begin{tabular}{ c c c c c c}
\hline \hline
     Star  &  HJD        & Error   & Type & Filter & Source    \\
           & 2400000+    & [days]  &      &        &           \\ \hline
 AM CrB    & 54248.47469 & 0.00384 & Sec   & W  &  Super WASP  \\
 AM CrB    & 54249.52519 & 0.00076 & Prim  & W  &  Super WASP  \\
 AM CrB    & 54250.57809 & 0.00277 & Sec   & W  &  Super WASP  \\
 AM CrB    & 54251.63576 & 0.00106 & Prim  & W  &  Super WASP  \\
 AM CrB    & 54252.34144 & 0.00013 & Prim  & W  &  Super WASP  \\
 AM CrB    & 54254.45026 & 0.00112 & Prim  & W  &  Super WASP  \\
\hline
\end{tabular}}
\end{table}

\begin{table}
 \tiny
 \caption{The heliocentric minima times used for the analysis.}
 \label{TableMin} \centering \scalebox{0.80}{
\begin{tabular}{ c c c c c c}
\hline \hline
     Star  &  HJD        & Error   & Type & Filter & Source    \\
           & 2400000+    & [days]  &      &        &           \\ \hline
 V1298Her  & 53236.43182 & 0.01652 & Sec   &  W  &  Super WASP  \\
 V1298Her  & 53237.50335 & 0.00197 & Prim  &  W  &  Super WASP  \\
 V1298Her  & 53238.57342 & 0.00756 & Sec   &  W  &  Super WASP  \\
 V1298Her  & 53239.65267 & 0.00609 & Prim  &  W  &  Super WASP  \\
 V1298Her  & 53248.18569 & 0.00724 & Prim  &  W  &  Super WASP  \\
 V1298Her  & 53249.30336 & 0.03392 & Sec   &  W  &  Super WASP  \\
\hline
\end{tabular}}
\end{table}

\begin{table}
 \tiny
 \caption{The heliocentric minima times used for the analysis.}
 \label{TableMin} \centering \scalebox{0.80}{
\begin{tabular}{ c c c c c c}
\hline \hline
     Star  &  HJD        & Error   & Type & Filter & Source    \\
           & 2400000+    & [days]  &      &        &           \\ \hline
 EL Lyn    & 54156.51633 & 0.00360 & Prim  &  W  &  Super WASP  \\
 EL Lyn    & 54157.45795 & 0.00277 & Sec   &  W  &  Super WASP  \\
 EL Lyn    & 54158.44848 & 0.00229 & Prim  &  W  &  Super WASP  \\
 EL Lyn    & 54163.59946 & 0.00222 & Prim  &  W  &  Super WASP  \\
 EL Lyn    & 54165.53879 & 0.00364 & Prim  &  W  &  Super WASP  \\
 EL Lyn    & 54166.51572 & 0.00299 & Sec   &  W  &  Super WASP  \\
\hline
\end{tabular}}
\end{table}

\begin{table}
 \tiny
 \caption{The heliocentric minima times used for the analysis.}
 \label{TableMin} \centering \scalebox{0.80}{
\begin{tabular}{ c c c c c c}
\hline \hline
     Star  &  HJD        & Error   & Type & Filter & Source    \\
           & 2400000+    & [days]  &      &        &           \\ \hline
 FW Per    & 54383.74596 & 0.00139 & Sec   &  W  &  Super WASP  \\
 FW Per    & 54384.54465 & 0.00246 & Sec   &  W  &  Super WASP  \\
 FW Per    & 54387.70520 & 0.00102 & Sec   &  W  &  Super WASP  \\
 FW Per    & 54388.50291 & 0.00352 & Sec   &  W  &  Super WASP  \\
 FW Per    & 54389.68444 & 0.00068 & Prim  &  W  &  Super WASP  \\
 FW Per    & 54392.44930 & 0.00318 & Sec   &  W  &  Super WASP  \\
\hline
\end{tabular}}
\end{table}

\begin{table}
 \tiny
 \caption{The heliocentric minima times used for the analysis.}
 \label{TableMin} \centering \scalebox{0.80}{
\begin{tabular}{ c c c c c c}
\hline \hline
     Star  &  HJD        & Error   & Type & Filter & Source    \\
           & 2400000+    & [days]  &      &        &           \\ \hline
 RU Tri    & 53975.57432 & 0.02329 &  Prim &  W  &  Super WASP  \\
 RU Tri    & 53978.86989 & 0.01711 &  Prim &  W  &  Super WASP  \\
 RU Tri    & 53980.40233 & 0.01423 &  Sec  &  W  &  Super WASP  \\
 RU Tri    & 53982.25314 & 0.01465 &  Prim &  W  &  Super WASP  \\
 RU Tri    & 53991.91971 & 0.00235 &  Prim &  W  &  Super WASP  \\
 RU Tri    & 53993.57382 & 0.00048 &  Sec  &  W  &  Super WASP  \\
\hline
\end{tabular}}
\end{table}

\begin{table}
 \tiny
 \caption{The heliocentric minima times used for the analysis.}
 \label{TableMin} \centering \scalebox{0.80}{
\begin{tabular}{ c c c c c c}
\hline \hline
     Star  &  HJD        & Error   & Type & Filter & Source    \\
           & 2400000+    & [days]  &      &        &           \\ \hline
 RU Tri    & 54436.50605 & 0.00303 &  Prim &  W  &  Super WASP  \\
 RU Tri    & 54438.08309 & 0.00911 &  Sec  &  W  &  Super WASP  \\
 RU Tri    & 54439.74973 & 0.00238 &  Prim &  W  &  Super WASP  \\
 RU Tri    & 54441.42159 & 0.00748 &  Sec  &  W  &  Super WASP  \\
 RU Tri    & 54444.66103 & 0.01172 &  Sec  &  W  &  Super WASP  \\
 RU Tri    & 54446.29983 & 0.00812 &  Prim &  W  &  Super WASP  \\
\hline

 \hline
\end{tabular}}
\end{table}
%
%
%
%
%


\begin{thebibliography}{}

 \bibitem[Akerlof et al.(2000)]{2000AJ....119.1901A} Akerlof, C., Amrose, S., Balsano, R., et al.\ 2000, \aj, 119, 1901
 \bibitem[Butters et al.(2010)]{2010A&A...520L..10B} Butters, O.~W., West, R.~G., Anderson, D.~R., et al.\ 2010, A\&A, 520, L10
 \bibitem[Guinan \& Engle(2006)]{2006Ap&SS.304....5G} Guinan, E.~F., \& Engle, S.~G.\ 2006, Ap\&SS, 304, 5
 \bibitem[Harmanec (1988)]{1988BAICz..39..329H} Harmanec, P.\ 1988, BAICz, 39, 329
 \bibitem[Hoffmeister(1943)]{1943AN....274...36H} Hoffmeister, C.\ 1943, AN, 274, 36
 \bibitem[Ho\v{n}kov\'a et al.(2013)]{OEJV160} Ho\v{n}kov\'a, K., et al.\ 2013, OEJV, 160, 1
 \bibitem[Khruslov(2008)]{2008PZP.....8...41K} Khruslov, A.~V.\ 2008, Peremennye Zvezdy Prilozhenie, 8, 41
 \bibitem[Lucy(1968)]{Lucy1968} Lucy, L.~B.\ 1968, \apj, 151, 1123
 \bibitem[Malkov~et~al.(2006)]{2006A&A...446..785M} Malkov, O.~Y., Oblak, E., Snegireva, E.~A., \& Torra, J.\ 2006, A\&A, 446, 785
 \bibitem[Mason et~al. (2001)]{WDS} Mason, B.~D., Wycoff, G.~L., Hartkopf, W.~I., Douglass, G.~G., \& Worley, C.~E. 2001, AJ, 122, 3466
 \bibitem[Norton et al.(2007)]{2007A&A...467..785N} Norton, A.~J., Wheatley, P.~J., West, R.~G., et al.\ 2007, A\&A, 467, 785
 \bibitem[Otero et al.(2005)]{2005IBVS.5586....1O} Otero, S.~A., Wils, P., \& Dubovsk\'y, P.~A.\ 2005, IBVS, 5586, 1
 \bibitem[Paschke \& Br\'at(2006)]{2006OEJV...23...13P} Paschke, A., \& Br\'at, L.\ 2006, OEJV, 23, 13
 \bibitem[Pickles \& Depagne(2010)]{2010PASP..122.1437P} Pickles, A., \& Depagne, {\'E}.\ 2010, PASP, 122, 1437
 \bibitem[Pojmanski(2002)]{2002AcA....52..397P} Pojmanski, G.\ 2002, AcA, 52, 397
 \bibitem[Pollacco et al.(2006)]{2006PASP..118.1407P} Pollacco, D.~L., et al.\ 2006, PASP, 118, 1407
 \bibitem[Pr{\v s}a \& Zwitter(2005)]{Prsa2005} Pr{\v s}a, A., Zwitter, T.\ 2005, \apj, 628, 42
 \bibitem[Roeser \& Bastian(1988)]{1988A&AS...74..449R} Roeser, S., \& Bastian, U.\ 1988, A\&AS, 74, 449
 \bibitem[Skiff(2007)]{2007yCat.2277....0S} Skiff, B.~A.\ 2007, VizieR Online Data Catalog, 2277, 0
 \bibitem[Samus et al. (2012)]{GCVS} Samus N.N., Durlevich O.V., Kazarovets E V., et al. General Catalog of Variable Stars (GCVS database, Version 2012Feb)
 \bibitem[Skrutskie et al.(2006)]{2006AJ....131.1163S} Skrutskie, M.~F., Cutri, R.~M., Stiening, R., et al.\ 2006, \aj, 131, 1163
 \bibitem[Strohmeier(1955)]{1955VeBam..11....1S} Strohmeier, W.\ 1955, VeBam, 11, 1
 \bibitem[van Hamme(1993)]{vanHamme1993} van Hamme, W.\ 1993, \aj, 106, 2096
 \bibitem[Weber(1963)]{1963IBVS...21....1W} Weber, R.\ 1963, IBVS, 21, 1
 \bibitem[Wo{\'z}niak et al.(2004)]{2004AJ....127.2436W} Wo{\'z}niak, P.~R., Vestrand, W.~T., Akerlof, C.~W., et al.\ 2004, \aj, 127, 2436
 \bibitem[Wilson \& Devinney(1971)]{Wilson1971} Wilson, R.~E., Devinney, E.~J.\ 1971, \apj, 166, 605
 \bibitem[Zasche et al.(2014)]{ZascheSMC2014} Zasche, P., Wolf, M., Vra\v{s}til, J., Li\v{s}ka, J., Skarka, M., Zejda, M.\ 2014, A\&A, accepted, DOI: 10.1051/0004-6361/201424273.

\end{thebibliography}
\end{document}